\documentclass[12pt]{iopart}

\expandafter\let\csname equation*\endcsname\relax
\expandafter\let\csname endequation*\endcsname\relax

\usepackage{iopams,mathrsfs,graphicx,epstopdf,bigints}

\DeclareMathAlphabet{\mathscrbf}{OMS}{mdugm}{b}{n}

\newcommand{\Lagr}{\mathscr{L}}

\begin{document}

\title[Magnetic monopoles and dyons revisited]{Magnetic monopoles and dyons revisited: A useful contribution to the study of classical mechanics}

\author{Renato P dos Santos}

\address{PPGECIM, ULBRA - Lutheran University of Brazil,
Av. Farroupilha, 8001 - Pr. 14, S. 338 - 92425-900 Canoas, RS, Brazil}

\ead{renatopsantos@ulbra.edu.br}

\begin{abstract}
Graduate level physics curricula in many countries around the world, as well as senior-level undergraduate ones in some major institutions, include Classical Mechanics courses, mostly based on Goldstein's textbook masterpiece. During the discussion of central force motion, however, the Kepler problem is virtually the only serious application presented. In this paper, we present another problem that is also soluble, namely the interaction of Schwinger's dual-charged (dyon) particles. While the electromagnetic interaction of magnetic monopoles and electric charges was studied in detail some 40 years ago, we consider that a pedagogical discussion of it from an essentially classical mechanics point of view is a useful contribution for students. Following a path that generalizes Kepler's problem and Rutherford scattering, we show that they exhibit remarkable properties such as stable non-planar orbits, as well as rainbow and glory scattering, which are not present in the ordinary scattering of two singly charged particles. Moreover, it can be extended further to the relativistic case and to a semi-classical quantization, which can also be included in the class discussion.
\end{abstract}
\pacs{0140gb, 4520Jj, 0350De, 1480Hv}

\noindent{\it Keywords\/}: Physics teaching, Classical Mechanics, Lagrangian mechanics, Electromagnetism, Dirac magnetic monopoles, dyons

\submitto{\EJP}

\maketitle

\section{Introduction}
In Classical Mechanics courses, the so-called Kepler problem~\cite[ch.~3, sec. 3.7]{GOLDSTEINhCla} is virtually the only serious completely integrable application of central forces discussed besides the harmonic oscillator. As Sivardi\`ere already pointed out in this journal~\cite{SIVARDIEREjCla}, the motion of a charged particle in the field of a magnetic monopole~\cite{DIRACpQua}, which is another example of completely integrable problem, is, unfortunately, not discussed. 

Here, we extend and deepen Sivardi\`ere's study to a more general case, namely the interaction between Schwinger's dyons (dual charged  particles)~\cite{SCHWINGERjMod}. As it has formal similarity with the Kepler problem and the Rutherford scattering, we believe that it may be presented right after these ones in Classical Mechanics courses. Furthermore, it exhibits unusual features such as non-planar stable orbits and rainbow and glory scattering, results that may arouse students' and teachers' interest.

%382p

\section{A brief historical review of magnetic monopoles and dyons}
The similarity between the electric and magnetic fields is visible in Maxwell's equations. When deducing them in 1873, Maxwell himself pointed out that it would be necessary \emph{to assert} that there are no net magnetic charged bodies and no `magnetic currents'~\cite[art.~380, p.~6]{MAXWELLjTre}.

However, this symmetry would be restored if we were to assume the existence of a magnetic field density~$\rho_{\mathrm{m}}$ and a `magnetic current'~$\bi{j}_{\mathrm{m}}$, obtaining (in SI units and with magnetic charges measured in ampere $\cdot$ meters)
\begin{eqnarray}
 \begin{cases}
	\bnabla \cdot \bi{E} &= \rho_{\mathrm{e}} / \epsilon_0 , \\
	\bnabla \times \bi{E} &= - \mu_0 \bi{j}_{\mathrm{m}} - \partial\bi{B} / \partial t , \\
	\bnabla \cdot \bi{B} &= \mu_0 \rho_{\mathrm{m}} , \\
	\bnabla \times \bi{B} &= \mu_0 \bi{j}_{\mathrm{e}} + \mu_0 \epsilon_0 \partial\bi{E} / \partial t .  
 \end{cases}
 \label{1.4}
\end{eqnarray}

In 1896, H. Poincar\'e, applied the concept of \emph{magnetic matter} to explain Birkeland's magnetic deflection of cathode rays experiment. In it, he considered that the beam passes so close to one pole of the magnet that the other can be neglected~\cite{POINCAREhRem}. This can be interpreted as a \emph{magnetic monopole} approximation.

Nevertheless, the idea of a magnetic monopole as a particle having a single magnetic pole was introduced only in 1931 by Dirac, in his famous work~\cite{DIRACpQua}.

As a matter of fact, Dirac was not looking for something like the monopole, but investigating the why of the quantization of electric charge. In other words, why the electric charge always appears in Nature as a multiple of the electron charge~$e$ and why this charge has a value such that~(in the same units as in~\eref{1.4})
\begin{eqnarray}
	2 \epsilon_0 h c/e^2 \cong 137 . \label{2.1}
\end{eqnarray}

In that work, however, instead of the relation~\eref{2.1}, he obtained
\begin{eqnarray}
	e g / \epsilon_0 h c^2 = n \qquad & (n = \pm1, \pm2, \pm3, \ldots ) , \label{2.10}
\end{eqnarray}
which is known as the \emph{Dirac quantization condition}. It does not set a value for~$e$, but only for its product by the magnetic charge~$g$ of a hypothetical particle in its vicinity. On the other hand, Caruso~\cite{CARUSOfOri} arrived at the same result~\eref{2.10} through a semi-classical derivation that gives it a new interpretation.

Despite that, in a way,~\eref{2.10} offers a solution to Dirac's initial problem: if there were a single magnetic monopole in the entire Universe, then all electric charges would be quantized according to~\eref{2.10}. Therefore, in view of the observed quantization of electric charge and in the absence of another explanation for this fact at the time, the condition~\eref{2.10} was considered a serious argument for the existence of at least one magnetic monopole in the Universe.

Generalizing Dirac quantization condition, Schwinger~\cite{SCHWINGERjMod}, in 1969, introduced the dyons, which were suggested at the time as candidates for the quark model. Pinfold et al.~\cite{PINFOLDjDir} discuss the tremendous implications that the discovery of magnetic monopoles or dyons would have for our fundamental understanding of Nature at the deepest levels and describe the search for these particles, from Dirac's proposal in 1931 to the current MoEDAL experiment at CERN's LHC.

After this short historical account, we proceed to study the interaction of two dyons from a classical mechanics point of view.

\section{The classical mechanics of two dyons}
To formulate the Lagrangian for the electromagnetic interaction of two dyons, we need to circumvent the problem of what became known as the \emph{Dirac string}~\cite{DIRACpQua}. It is a singularity line in the space starting from the monopole, over which the magnetic vector potential~$\bi{A}$ does not satisfy the condition that the wave function is a univalent function. 

Later on, Schwinger~\cite{SCHWINGERjCha}, Yang~\cite{YANGcMag}, and Wu and Yang~\cite{WUtDir} considered that the Dirac string has no physical meaning or real existence, being only an effect of the coordinate system, analogous to the problem of terrestrial geographic poles when trying to map the Earth's surface with a single chart. These authors stated that this problem can be solved by dividing the space around the monopole in two regions~$a$ and~$b$ and defining two potential vectors~$\bi{A}_a$ and~$\bi{A}_b$ which describe the monopole field in each of these regions and have singularities in the other~$b$ and~$a$ ones, respectively.

Bollini and Giambiagi~\cite{BOLLINIcGau}, in their turn, proposed a multivalued distribution (generalized function) potential instead of a singular function. This approach, however, demands the use of the mathematical theory of distributions to the evaluation of its curls and divergences.

On the other hand, Sokolov~\cite{SOKOLOVvSch} showed that the singularity of the magnetic monopole potential is of a purely kinematic origin, caused by the uncertainty of the azimuthal~$\phi$ angle along the~$z$ axis. As a consequence, for a charge in the field of a 'Coulombian' magnetic monopole 
\begin{eqnarray}
 \bi{B} = \frac{\mu_0}{4 \pi} \frac{g}{r^3}\bi{r} \label{4.1}
\end{eqnarray}
we can obtain the equation of motion, by using the Lorentz force 
\begin{eqnarray}
  \bi{F} &= e \bi{v} \times \bi{B} 
\end{eqnarray}
and the expression~\eref{4.1} for the magnetic field, as 
\begin{eqnarray}
 \mu\frac{\rmd \bi{v}}{\rmd t} = \frac{\mu_0}{4 \pi} eg \bi{v} \times \frac{\bi{r}}{r^3} \label{4.19}
\end{eqnarray}
without the appearance of strings or fictitious fields as long as vector potentials and curls are written and evaluated in spherical coordinates in
\begin{eqnarray}
 \mu\frac{\rmd \bi{v}}{\rmd t} = e \bi{v} \times \left( \bnabla \times \bi{A} \right)_{\mathrm{spherical}} . \label{4.22}
\end{eqnarray}

Sokolov's procedure can be as well generalized to the dyon-dyon case~\cite[p.~13]{dosSANTOSrMon} for any vector potential $\bi{A}$ whose curl evaluated in spherical coordinates furnishes the correct magnetic potential~\eref{4.1} such that~\eref{4.22} leads to~\eref{4.19}. As a result, it is possible to obtain a classical string-free Lagrangian for the dyon-dyon case. 

That being said, we will here build the dyon-dyon Lagrangian through a different procedure, by means of the fields instead of a potential. This procedure will lead, however, to a Lagrangian that matches Sokolov's one.

We start by noticing that Maxwell's equations~\eref{1.4} are invariant under the duality transformation~$\bi{E} \to c\bi{B}$,~$c\bi{B} \to -\bi{E}$,~$\rho_\mathrm{e} \to \rho_\mathrm{m}/c$, and~$\bi{j}_\mathrm{e} \to \bi{j}_\mathrm{m}/c$. That allows us to generalize Lorentz force to the dyon-dyon interaction and write the equation of motion as
\begin{eqnarray}
 \mu\frac{\rmd \bi{v}}{\rmd t} &= e_1\left(\bi{E} + \bi{v} \times \bi{B} \right) + g_1\left(\bi{B} - \frac{1}{c^2}\bi{v} \times \bi{E} \right) \nonumber \\
  &= \left( e_1 \frac{1}{4 \pi \epsilon_0 } e_2 + g_1 \frac{\mu_0 }{4 \pi } g_2 \right) \frac{\bi{r}}{r^3} +  \left(e_1 \frac{\mu_0}{4 \pi} g_2 - g_1 \frac{1}{c^2} \frac{1}{4 \pi \epsilon_0} e_2 \right) \bi{v} \times \frac{\bi{r}}{r^3} \nonumber \\
  &= \frac{1}{4 \pi \epsilon_0 } \left( e_1 e_2 + \frac{1}{c^2} g_1 g_2 \right) \frac{\bi{r}}{r^3} + \frac{\mu_0}{4 \pi} \left(e_1 g_2 - g_1 e_2 \right) \bi{v} \times \frac{\bi{r}}{r^3} \nonumber \\
  &= \frac{1}{4 \pi \epsilon_0 } q \frac{\bi{r}}{r^3} - \frac{\mu_0}{4 \pi} \kappa \bi{v} \times \frac{\bi{r}}{r^3} \label{5.1}
\end{eqnarray}
where
\begin{eqnarray}
 \eqalign{
	q = e_1 e_2 + g_1 g_2 / c^2 \\
	\kappa = e_1 g_2 - g_1 e_2 , 
 }
\label{5.2}
\end{eqnarray}
being~$e_1$,~$e_2$,~$g_1$, and~$g_2$ the electric and magnetic charges of the two dyons, corresponding the index~$_2$ to the dyon that remains at the origin of the relative coordinate system,~$c$ the speed of light in a vacuum, and~$\mu$, naturally, the system \emph{reduced mass}, given by 
\begin{eqnarray}
	\mu = \frac{m_1 m_2}{m_1+m_2} .
\end{eqnarray}

\begin{figure}[ht]
    \centering
    \includegraphics[width=0.8\columnwidth]{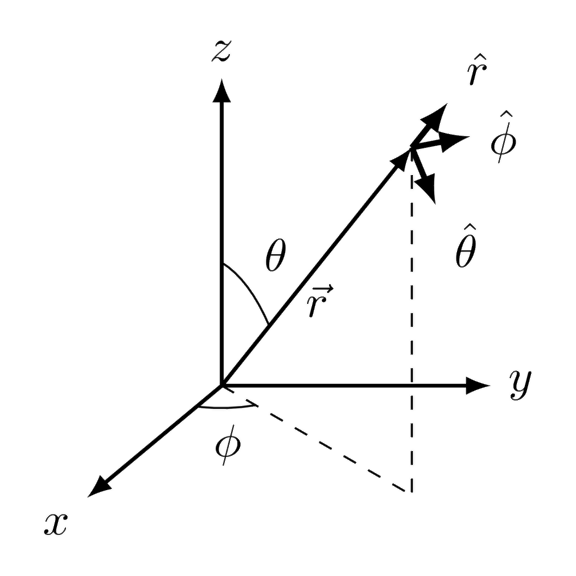}
    \caption{Position vector ($\bi{r}$) in the spherical coordinate system used in this paper.}
    \label{fig_1}
\end{figure}

To evaluate the cross product in~\eref{5.1}, we need to express~$\bi{r}$ and~$\bi{v}$ in spherical coordinates as (see \Fref{fig_1})
\begin{eqnarray}
 \eqalign{
  \bi{r} = r \hat{\bi{r}} \\
  \bi{v} = \dot{r} \hat{\bi{r}} + r \dot{\theta} \hat{\btheta} + r \sin\theta\dot{\phi} \:\hat{\bphi} . 
 } \label{vsphcoor}
 \end{eqnarray}

Now, we can write the force on the right-hand side of~\eref{5.1} as
\begin{eqnarray}
 \eqalign{
  \bi{F} &= \frac{1}{4 \pi \epsilon_0 } \frac{q}{r^2}\:\hat{\bi{r}} - \frac{\mu_0}{4 \pi} \left( \frac{\kappa}{r} \sin\theta\dot{\phi}\: \hat{\btheta} + \frac{\kappa}{r}\: \dot{\theta}\hat{\bphi} \right) \\
  &= \frac{1}{4 \pi \epsilon_0 } \frac{q}{r^2}\:\hat{\bi{r}} + \mathscrbf{F} , 
 }
 \label{5.3}
\end{eqnarray}
where~$\mathscrbf{F}$ includes the non-central terms of~$\bi{F}$ (those that not depend only on the distance $r$  and are not directed along the $\hat{\bi{r}}$ direction).

From~\eref{vsphcoor} and~\eref{5.3}, we can write the Lagrangian 
\begin{eqnarray}
	\Lagr &= T - V \nonumber \\
	&= \frac{1}{2}\mu\dot{r}^2 + \frac{1}{2}\mu r^2\dot{\theta}^2 + \frac{1}{2}\mu r^2\sin^2\theta\:\dot{\phi}^2 - \frac{1}{4 \pi \epsilon_0 } \frac{q}{r} , \label{5.4}
\end{eqnarray}
which we call `incomplete' because it contains only the term in~\eref{5.3} that is derivable from a scalar (Coulombian) potential.

As not all the forces acting on the system are derivable from a scalar potential, then Lagrange's equations can be written in the inhomogeneous form~\cite[sec.~1.5]{GOLDSTEINhCla} 
\begin{eqnarray}
	\frac{\rmd }{\rmd t}\left(\frac{\partial\Lagr}{\partial\dot{q}_i} \right) - \frac{\partial\Lagr}{\partial q_i} = Q_i \qquad & (i=1,2,3,\cdots) ,  \label{5.5}
\end{eqnarray}
where~$Q_i$ are the \emph{generalized forces} defined by
\begin{eqnarray}
 Q_i = \mathscrbf{F}_i \cdot \frac{\partial\bi{r}}{\partial q_i} .  \label{5.6}
\end{eqnarray}

Substituting~\eref{5.4}, \eref{5.3} and~\eref{5.6} into \eref{5.5}, we obtain the equations of motion as
\begin{eqnarray}
 \eqalign{
	\mu\ddot{r}-\mu r\dot{\theta}^2 - \mu r\sin^2\theta\:\dot{\phi}^2 - \frac{1}{4 \pi \epsilon_0 } \frac{q}{r^2} = 0  \\
	\mu r^2\ddot{\theta} + 2\mu r\dot{r}\dot{\theta} - \mu r^2\sin\theta\cos\theta\:\dot{\phi}^2 = -\frac{\mu_0}{4 \pi} \kappa\sin\theta\:\dot{\phi} \\
	\mu r^2\sin^2\theta\:\ddot{\phi} + 2\mu r^2\sin\theta\cos\theta\:\dot{\theta}\dot{\phi} + 2\mu r\dot{r}\sin^2\theta\:\dot{\phi}^2 = \frac{\mu_0}{4 \pi} \kappa\sin\theta\:\dot{\theta} . 
 } 
 \label{5.7}
\end{eqnarray}

It is worthy of note, however, that the equations of motion~\eref{5.7} could also have been obtained from a Lagrangian of the form
\begin{eqnarray}
	\Lagr = \frac{1}{2}\mu\dot{r}^2 + \frac{1}{2}\mu r^2\dot{\theta}^2 + \frac{1}{2}\mu r^2\sin^2\theta\:\dot{\phi}^2 - \frac{1}{4 \pi \epsilon_0 } \frac{q}{r}  + \frac{\mu_0}{4 \pi} \kappa\cos\theta\dot{\phi} \label{5.8}
\end{eqnarray}
which we call the `minimal' Lagrangian as it is the simpler one that furnishes the equations~\eref{5.7} that describe the classical interaction of two dyons without the appearance of Dirac strings, according to our generalization of Sokolov's procedure.

Now, it is important to note that, since the interaction force~\eref{5.3} is not central, we should not expect the vector mechanical angular momentum~$\bi{L}$ to be conserved. As a matter of fact, from its expression in spherical coordinates, using~\eref{vsphcoor} again,
\begin{eqnarray}
 \eqalign{
	\bi{L} &= \mu\bi{r}\times\bi{v} \\
	&= \mu r^2\sin\theta\:\dot{\phi}\:\hat{\btheta} + \mu r^2\dot{\theta}\:\hat{\bphi} , 
 }
	\label{6.1}
\end{eqnarray}
we can evaluate its temporal derivative by remembering that
\begin{eqnarray}
 \eqalign{
 \frac{\rmd \hat{\bi{r}}}{\rmd t} = \dot{\theta}\hat{\btheta} + \sin\theta\dot{\phi}\:\hat{\bphi}  \\ 
 \frac{\rmd \hat{\btheta}}{\rmd t} = -\dot{\theta}\hat{\bi{r}} + \cos\theta\dot{\phi}\:\hat{\bphi} \\
 \frac{\rmd \hat{\bphi}}{\rmd t} = -\sin\theta\dot{\phi}\:\hat{\bi{r}} - \cos\theta\dot{\phi}\:\hat{\bphi} , 
 }
\end{eqnarray}
obtaining traightforwardly
\begin{eqnarray}
 \eqalign{
  \left(\frac{\rmd \bi{L}}{\rmd t}\right)_r = 0 , \\
	\left(\frac{\rmd \bi{L}}{\rmd t}\right)_\theta = -\mu r^2\sin\theta\:\ddot{\phi} - 2\mu r^2\cos\theta\:\dot{\theta}\dot{\phi} - 2\mu r\dot{r}\sin\theta\:\dot{\phi}^2 \ \ \  \text{, and} \\
  \left(\frac{\rmd \bi{L}}{\rmd t}\right)_\phi = \mu r^2\ddot{\theta} + 2\mu r\dot{r}\dot{\theta} - \mu r^2\sin\theta\cos\theta\:\dot{\phi}^2  
	}
\end{eqnarray}

Comparing these results with the left-hand sides of the two last equations of motion~\eref{5.7} above, we conclude that
\begin{eqnarray}
	\left(\frac{\rmd \bi{L}}{\rmd t}\right)_\theta = - \frac{\mu_0}{4 \pi} \kappa\:\dot{\theta} \nonumber \ \ \  \text{and} \\
	\left(\frac{\rmd \bi{L}}{\rmd t}\right)_\phi = -\frac{\mu_0}{4 \pi} \kappa\sin\theta\:\dot{\phi} \nonumber  
\end{eqnarray}
or
\begin{eqnarray}
	\frac{\rmd \bi{L}}{\rmd t} = - \frac{\mu_0}{4 \pi} \kappa \frac{\rmd \hat{\bi{r}}}{\rmd t} \label{6.2}
\end{eqnarray}
and that the vector
\begin{eqnarray}
	\bi{J} \equiv \bi{L} + \frac{\mu_0}{4 \pi} \kappa\hat{\bi{r}} \label{6.3}
\end{eqnarray}
is conserved. 

The vector $\bi{J}$, defined by \eref{6.3}, is known as the \emph{Poincar\'e integral of motion}, as it was first found by Poincar\'e in his previously mentioned work~\cite{POINCAREhRem}. It can be interpreted as the `total' angular momentum of the system because the second term in the right-hand side of~\eref{6.3} is the angular momentum of the electromagnetic field, as demonstrated by Thomson~\cite[p.~532]{THOMSONjEle}.

On the other hand, from the definition~\eref{6.3} and the fact that~$\bi{L}$ is perpendicular to~$\hat{\bi{r}}$, we also obtain
\begin{eqnarray}
	J^2 = L^2 + \left( \frac{\mu_0}{4 \pi} \right)^2 \kappa^2 \label{6.4}
\end{eqnarray}
and, since~$\bi{J}$ is conserved and~$\kappa$ is a constant, we conclude that the module~$L$ of the angular momentum is conserved even though the vector~$\bi{L}$ is not.

We can now use the remaining equation of motion~\eref{5.7} to obtain the conservation of the total energy of the system. To do so, it can be rewritten, in terms of~$\bi{L}$ given by~\eref{6.1}, as
\begin{eqnarray}
	\mu\ddot{r}-\frac{L^2}{\mu r^3} - \frac{1}{4 \pi \epsilon_0 } \frac{q}{r^2} = 0 . \nonumber
\end{eqnarray}

Now, to proceed further, we may do the trick~\cite[p. 74]{GOLDSTEINhCla} of rewriting its right-hand side as a derivative in $r$ and multiplying both sides by $\dot{r}$ as
\begin{eqnarray}
	\mu\ddot{r}\dot{r} = -\frac{\rmd}{\rmd r}\left( \frac{L^2}{2\mu r^2} + \frac{1}{4 \pi \epsilon_0 } \frac{q}{r} \right) \dot{r} , \nonumber
\end{eqnarray}
from what, remembering that $\rmd f(r) / \rmd t= (\rmd r/ \rmd t)\rmd f(r) / \rmd r = \dot{r} \rmd f(r) / \rmd r$  and that $\rmd \dot{r}^2/ \rmd t = 2 \ddot{r}\dot{r}$, it follows that
\begin{eqnarray}
	\frac{\rmd}{\rmd t} \left( \frac{\mu}{2}\dot{r}^2  \right) = -\frac{\rmd}{\rmd t}\left( \frac{L^2}{2\mu r^2} + \frac{1}{4 \pi \epsilon_0 } \frac{q}{r} \right) , \nonumber
\end{eqnarray}
which expresses  the conservation of the total energy of the system
\begin{eqnarray}
	E = \frac{\mu}{2}\dot{r}^2 + \frac{L^2}{2\mu r^2} + \frac{1}{4 \pi \epsilon_0 } \frac{q}{r}. \label{6.5}
\end{eqnarray}

Notice, now, that the definitions~\eref{6.3} of the~$\bi{J}$ vector and~\eref{6.1} of the angular momentum~$\bi{L}$ lead to the result
\begin{eqnarray}
	\bi{J}\cdot\hat{\bi{r}} &= \left( \mu\bi{r}\times\bi{v} + \frac{\mu_0}{4 \pi} \kappa\hat{\bi{r}} \right) \cdot\hat{\bi{r}} \nonumber \\
	&= \frac{\mu_0}{4 \pi} \kappa \label{6.6}
\end{eqnarray}
where we used the fact that $\bi{r}\times\bi{v} \cdot\hat{\bi{r}} = 0$.

If we interpret $\bi{J}\cdot\hat{\bi{r}}$ as the projection of $\bi{J}$ on the direction of $\hat{\bi{r}}$, we can define~$\alpha$ as the angle formed by~$\hat{\bi{r}}$ and~$\bi{J}$ given by
\begin{eqnarray}
	\cos\alpha \equiv \frac{\bi{J}\cdot\hat{\bi{r}}}{J} \nonumber
\end{eqnarray}
which, as we see from~\eref{6.6} and~\eref{6.3}, is constant, with value
\begin{eqnarray}
	\alpha = \arccos( \mu_0 \kappa / 4 \pi J ) \label{6.7} 
\end{eqnarray}
or, by using the trigonometric identity
\begin{eqnarray}
 \tan(\arccos x) = \frac{\sqrt{1 - x^2}}{x} , \nonumber
\end{eqnarray}
we get from~\eref{6.7} and~\eref{6.4} 
\begin{eqnarray}
  \alpha &= \arctan \left( \frac{\sqrt{1 - ( \mu_0 \kappa / 4 \pi J )^2} }{ \mu_0 \kappa / 4 \pi J } \right) \nonumber \\
  &= \arctan \left( \frac{4 \pi J\sqrt{1 - ( \mu_0 \kappa / 4 \pi J )^2} }{ \mu_0 \kappa } \right) \nonumber \\
  &= \arctan \left( \frac{4 \pi \sqrt{ J^2 - ( \mu_0 / 4 \pi )^2 \kappa^2 } }{ \mu_0 \kappa }  \right) \nonumber \\
	&= \arctan( 4 \pi \sqrt{ L^2 } / \mu_0 \kappa ) \nonumber \\
	&= \arctan( 4 \pi L / \mu_0 \kappa ). \label{6.8}
\end{eqnarray}

\begin{figure}[ht]
    \centering
    \includegraphics[width=0.8\columnwidth]{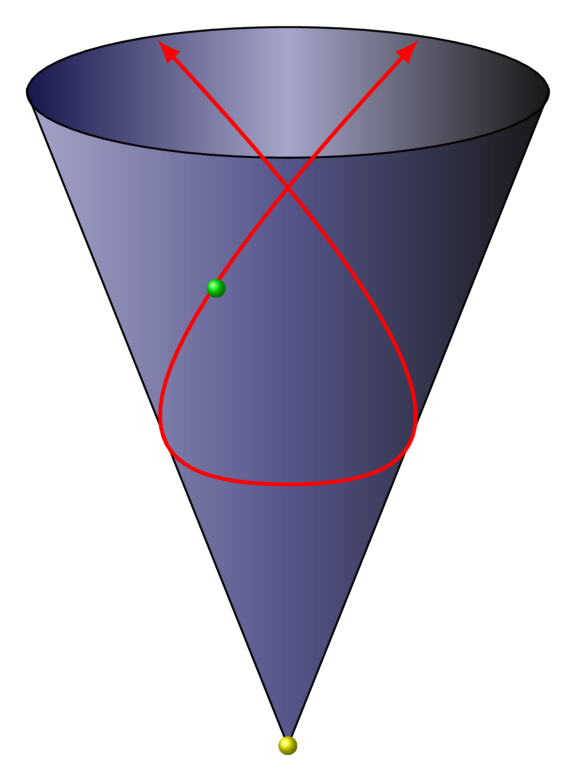}
    \caption{The orbit of the dyon confined to the surface of the Poincar\'e cone with half-aperture angle~$\alpha$, according to~\eref{8.9}.}
    \label{fig_2}
\end{figure}

Now, being~$\bi{J}$ a vector fixed in the space and the angle it forms with~$\hat{\bi{r}}$ constant, it implies that the motion is limited to the surface of a cone (the \emph{Poincar\'e cone}) (\Fref{fig_2}) of constant half-aperture angle~$\alpha$ given by
\begin{eqnarray}
	\alpha = \arccos( \mu_0 |\kappa| / 4 \pi J ), \label{6.9}
\end{eqnarray}
with~$\bi{J}$ being coincident with the interior axis if~$\kappa$ is positive and with its exterior axis if~$\kappa$ is negative. A similar conclusion was obtained by Poincar\'e~\cite{POINCAREhRem} for the movement of an electric charge in the field of a pole of a magnet (equivalent to a Dirac's monopole) and Appel~\cite{APPELpMou} for the movement of a electric charge in the field of a magnetic and electric pole simultaneously (equivalent to a Schwinger's dyon).

The fact that the motion is limited to the surface of a cone of constant half-aperture angle~$\alpha$  allows us to choose a new spherical coordinate system~($r$, $\alpha$, $\beta$), in which the vector~$\bi{J}$ coincides with the polar axis, so as to have only the two degrees of freedom radial distance~($r$) and azimuthal angle~($\beta$).

With this coordinate system,~$\dot{\alpha}=0$, while the Lagrangian~\eref{5.8} reduces to
\begin{eqnarray}
	\Lagr = \frac{1}{2}\mu\dot{r}^2 + \frac{1}{2}\mu r^2\sin^2\alpha\:\dot{\beta}^2 - \frac{1}{4 \pi \epsilon_0 } \frac{q}{r} + \frac{\mu_0}{4 \pi} \kappa\cos\alpha\:\dot{\beta}	\label{6.12}
\end{eqnarray}
and the equations of motion~\eref{5.7} result
\begin{eqnarray}
 \eqalign{
	\mu\ddot{r} - \mu r\sin^2\alpha\:\dot{\beta}^2 - \frac{1}{4 \pi \epsilon_0 } \frac{q}{r^2} = 0 \\
	\mu r^2\sin\alpha\cos\alpha\:\dot{\beta}^2 = \frac{\mu_0}{4 \pi} \kappa\sin\alpha\:\dot{\beta} \\
	2\mu r\dot{r}\sin^2\alpha\:\dot{\beta}^2 + \mu r^2\sin^2\alpha\:\ddot{\beta} = 0 . 
	}
	\label{6.13}
\end{eqnarray}

In the same way, the angular momentum~\eref{6.1} is now expressed by
\begin{eqnarray}
	\bi{L} = -\mu r^2\sin\alpha\:\dot{\beta}\hat{\balpha} ,
\end{eqnarray}
the conserved module of the angular momentum as 
\begin{eqnarray}
	L = \mu r^2\sin\alpha\:\dot{\beta} , \label{6.15}
\end{eqnarray}
and the conserved energy~\eref{6.5} results
\begin{eqnarray}
	E = \frac{\mu\dot{r}^2}{2} + \frac{L^2}{2\mu r^2} + \frac{1}{4 \pi \epsilon_0 } \frac{q}{r} . \label{6.14}
\end{eqnarray}

Now, to arrive at the equation of the orbit, we will follow Goldstein's procedure~\cite[sec.~3.7]{GOLDSTEINhCla}.

To start with, we can solve~\eref{6.14} for~$\dot{r}$ and get
\begin{eqnarray}
 \dot{r} = \sqrt{ \frac{2}{\mu}\left(E - \frac{1}{4 \pi \epsilon_0 } \frac{q}{r} - \frac{L^2}{2\mu r^2} \right) } \label{G3.16}
\end{eqnarray}
or, being $\dot{r} = \rmd r / \rmd t$,
\begin{eqnarray}
 \rmd t = \left(\frac{2E}{\mu} - \frac{q}{2 \pi \epsilon_0 \mu r} - \frac{L^2}{\mu^2 r^2} \right)^{-1/2} \rmd r . \label{G3.17}
\end{eqnarray}

For the equation of the orbit, we need the dependence of~$r$ upon~$\theta$ eliminating the parameter~$t$. This elimination can be done by seeing~\eref{6.15} as a relation between~$\rmd \beta$ and~$\rmd t$, in the same way as we did for~$\dot{r}$:
\begin{eqnarray}
	L\rmd t = \mu r^2\sin\alpha\:\rmd \beta \label{G3.31}
\end{eqnarray}
or
\begin{eqnarray}
	\rmd \beta = \frac{L}{\mu r^2\sin\alpha} \rmd t . \label{G3.32}
\end{eqnarray}

The substitution of~\eref{G3.17} into~\eref{G3.32} yields
\begin{eqnarray}
 \rmd \beta = \frac{L}{\mu r^2\sin\alpha}\left(\frac{2E}{\mu} - \frac{q}{2 \pi \epsilon_0 \mu r} - \frac{L^2}{\mu^2 r^2} \right)^{-1/2} \rmd r . \label{G3.35}
\end{eqnarray}

Now, integrating~\eref{G3.35} after slight rearrangements, we obtain
\begin{eqnarray}
 \beta = \frac{1}{\sin\alpha} \int \left( \frac{2\mu E}{L^2} - \frac{\mu q}{2 \pi \epsilon_0 L^2 r} - \frac{1}{r^2} \right)^{-1/2} \frac{\rmd r}{r^2} + \beta' , \label{G3.36}
\end{eqnarray}
where~$\beta'$ is a constant of integration determined by the initial conditions and not necessarily being the same as the initial angle~$\beta_0$ at time~$t=0$.

Finally, changing the variable of integration to~$u=1/r$, we obtains
\begin{eqnarray}
 \beta = \beta' - \frac{1}{\sin\alpha} \int \! \left(\frac{2\mu E}{L^2} - \frac{\mu q}{2 \pi \epsilon_0 L^2} u - u^2 \right)^{-1/2} \rmd u . \label{G3.50}
\end{eqnarray}

This indefinite integral is of the standard form~\cite[p.~93]{GOLDSTEINhCla}
\begin{eqnarray}
 \int \frac{\rmd x}{\sqrt{a + b u + c u^2}} = \frac{1}{\sqrt{-c}}\arccos\left(-\frac{b + 2c u}{\sqrt{\Delta}} \right) , \label{G3.51}
\end{eqnarray}
where
\begin{eqnarray}
 \Delta = b^2 - 4ac ,  \label{G3.51b}
\end{eqnarray}
where we identify
\begin{eqnarray}
 \eqalign{
a = 2 \mu E / L^2 , \\
b = -\mu q / 2 \pi \epsilon_0 L^2 \ \ \  \text{, and}\\
c = -1 .
 } 
 \label{G3.52}
\end{eqnarray}

Applying~\eref{G3.52} to~\eref{G3.51b}, we obtain
\begin{eqnarray}
 \Delta &= b^2(1 - 4ac/b^2) \nonumber \\
 &= \left( \frac{\mu q}{2 \pi \epsilon_0 L^2 } \right)^2 \left[ 1 + 4 \frac{2 \mu E}{ L^2 }\left( \frac{2 \pi \epsilon_0 L^2 }{\mu q} \right)^2 \right] \nonumber \\
 &= \left( \frac{\mu q}{2 \pi \epsilon_0 L^2 } \right)^2 \left[ 1 + \frac{2(4 \pi \epsilon_0 )^2 E L^2}{\mu q^2} \right] 
\end{eqnarray}
while applying~\eref{G3.52} to~$b + 2c x$ results
\begin{eqnarray}
	b + 2c x &= b(1+2c/bx) \nonumber \\
	&= \left( \frac{\mu q}{2 \pi \epsilon_0 L^2 } \right) \left( 1-2\frac{2 \pi \epsilon_0 L^2 }{\mu q}u \right) .
\end{eqnarray}
 
Now inserting these results into~\eref{G3.50}, we obtains
\begin{eqnarray}
 \beta = \beta' - \frac{1}{\sin\alpha} \arccos \left[ \left( \frac{4 \pi \epsilon_0 L^2u}{\mu q} - 1 \right) \Bigg/ \sqrt{ 1 + \frac{2(4 \pi \epsilon_0 )^2 E L^2}{\mu q^2} } \right] \label{G3.54}
\end{eqnarray}
which we can solve for~$u=1/r$, obtaining the equation of the orbit as
\begin{eqnarray}
	\frac{1}{r} = -\frac{\mu q}{4 \pi \epsilon_0 L^2}\left\{ 1 + \sqrt{1 + \frac{32 \pi^2 \epsilon_0^2 E L^2}{\mu q^2}}\cos\left[ \sin\alpha\left( \beta - \beta' \right) \right] \right\} , \label{6.18}
\end{eqnarray}
where we now identify~$\beta'$ as one of the turning angles of the orbit. 

One sees that, except for the~$\sin\alpha$ term,~\eref{6.18} is very similar to the equation for the Kepler problem of the planetary orbits~\cite[p.~93]{GOLDSTEINhCla}. As a matter of fact, if the cone has been degenerated to a plane~($\alpha = \pi/2$), the equation of the orbit~\eref{6.18} results
\begin{eqnarray}
	\frac{1}{r} = -\frac{\mu q}{4 \pi \epsilon_0 L^2}\left[1 + \varepsilon\cos\left( \beta - \beta' \right) \right] , \label{6.19}
\end{eqnarray}
which represents a conic curve with eccentricity
\begin{eqnarray}
	\varepsilon = \sqrt{ 1 + 32 \pi^2 \epsilon_0^2 E L^2 / \mu q^2 } . \label{6.20}
\end{eqnarray}

Now, remembering that the distance between two points that are differentially separated on the surface of a cone with a half angle~$\alpha$ is
\begin{eqnarray}
	(\rmd s)^2 = (\rmd r)^2 + r^2(\sin\alpha\:\rmd \beta)^2 , \label{6.21}
\end{eqnarray}
while the distance between two points on the plane is given by 
\begin{eqnarray}
	(\rmd s)^2 = (\rmd r)^2 + r^2(\rmd \phi)^2 ,\label{6.22}
\end{eqnarray}
we interpret~\eref{6.18} as representing a conic-shaped orbit confined to the surface of the Poincar\'e cone, as shown in \Fref{fig_2}.

As in Kepler problem, such conic-shaped orbits divide themselves into bound states with elliptic-like orbits ($\varepsilon<1$) and scatterings with hyperbolic-like orbits ($\varepsilon>1$)~\cite[p.~94]{GOLDSTEINhCla}. We will explore both cases in the next two sections.

\section{Bound dyon-dyon states}
Let us consider first the bound states~($\varepsilon < 1$) of the dyon-dyon system.

From~\eref{6.20}, this corresponds to
\begin{eqnarray}
	\sqrt{ 1 + 32 \pi^2 \epsilon_0^2 E L^2 / \mu q^2 } < 1 \nonumber
\end{eqnarray}
or
\begin{eqnarray}
	1 + 32 \pi^2 \epsilon_0^2 E L^2 / \mu q^2 < 1 \nonumber \\
	32 \pi^2 \epsilon_0^2 E L^2 / \mu q^2 < 0 \nonumber
\end{eqnarray}
which, being everything else positive, implies
\begin{eqnarray}
	E < 0 , \label{7.4}
\end{eqnarray}
as expected.

Now, from~\eref{6.14},~\eref{7.4} implies
\begin{eqnarray}
	\frac{\mu\dot{r}^2}{2} + \frac{L^2}{2\mu r^2} + \frac{1}{4 \pi \epsilon_0 } \frac{q}{r} < 0 \nonumber \\
	\frac{1}{4 \pi \epsilon_0 } \frac{q}{r} < \frac{\mu\dot{r}^2}{2} + \frac{L^2}{2\mu r^2} \nonumber 
\end{eqnarray}
which again, being everything else positive, implies
\begin{eqnarray}
	q < 0 , \label{7.5}
\end{eqnarray}
i.e., an attractive interaction with
\begin{eqnarray}
	\frac{1}{4 \pi \epsilon_0 } \frac{|q|}{r} > \frac{\mu\dot{r}^2}{2} + \frac{L^2}{2\mu r^2} . \nonumber 
\end{eqnarray}

Being elliptic-like orbits, we can calculate, from~\eref{6.18}, its two turning points (points of least or greatest distance of one dyon from the other),~$r_{min}$ and~$r_{max}$, as
\begin{eqnarray}
 \eqalign{
  r_{min} = \frac{4 \pi \epsilon_0 L^2 }{ \mu|q|} (1+\varepsilon) \\
  r_{max} = \frac{4 \pi \epsilon_0 L^2 }{ \mu|q|} (1-\varepsilon) 
 } ,
 \label{7.7}
\end{eqnarray}
from which we can obtain the semi-major axis~$a$ and semi-minor axis~$b$ as
\begin{eqnarray}
  a &= \frac{r_{min} + r_{max}}{2}  \nonumber \\
	b &= \sqrt{a^2(1 - \varepsilon^2)} . \nonumber
\end{eqnarray}

However, it is more useful to have $a$ and $b$ expressions in terms of the energy of the system. To start with, we make use of the fact that the radial velocity $\dot{r}$ is zero at those turning points. Therefore, from~\eref{6.14}, the conserved energy at those points becomes
\begin{eqnarray}
	E = \frac{L^2}{2\mu r^2} + \frac{1}{4 \pi \epsilon_0 } \frac{q}{r} \nonumber
\end{eqnarray}
which can be rewritten in the form of a quadratic equation as 
\begin{eqnarray}
	r^2 - \frac{1}{4 \pi \epsilon_0 } \frac{q}{E}r - \frac{L^2}{2\mu E} = 0 \nonumber
\end{eqnarray}
having the turning points $r_{min}$ and $r_{max}$ as its roots. 
Now, it is well known that the sum of the roots of a quadratic equation equals the negative of the coefficient of its linear term. Therefore,
\begin{eqnarray}
  r_{min} + r_{max} = \frac{1}{4 \pi \epsilon_0 } \frac{q}{E} ,  \nonumber 
\end{eqnarray}
and we obtain the semi-major axis~$a$ as
\begin{eqnarray}
  a &= \frac{r_{min} + r_{max}}{2} = \frac{1}{8 \pi \epsilon_0 } \frac{q}{E} 
\end{eqnarray}
and the semi-minor axis~$b$ as
\begin{eqnarray}
  b &= \sqrt{a^2(1 - \varepsilon^2)} \nonumber \\
	&= \sqrt{ \left( \frac{1}{8 \pi \epsilon_0 } \frac{q}{E} \right)^2 \left[ 1 - \left( 1 + \frac{32 \pi^2 \epsilon_0^2 E L^2}{\mu q^2} \right) \right] } \nonumber \\
	&= \sqrt{ \left( \frac{q^2}{64 \pi^2 \epsilon_0^2 E^2} \right) \left( \frac{32 \pi^2 \epsilon_0^2 E L^2}{\mu q^2} \right) } \nonumber \\
	&= \frac{L}{\sqrt{2\mu|E|} } .
 \label{7.6}
\end{eqnarray}

Notice that the existence of~$r_{min}$ and~$r_{max}$ does not necessarily mean that the orbit is closed but only that it is 'bounded' by those limiting distances. 

To analyse the closedness of the orbits, we have to see that the term $\cos\left[ \sin\alpha\left( \beta - \beta' \right) \right]$ in \eref{6.18} implies that the ratio between the periodicities of the radial coordinate~($r$) and  the azimuthal angle~($\beta$) is given by~$\sin\alpha$. Consequently, it is this parameter that will determine if the orbit is closed or not. 

To understand that, we have to consider that, at each revolution, the dyon describes a portion~$\sin\alpha$ of the ellipse. Now, the orbit will be closed if, say, after~$n$ revolutions, it will have completed exactly~$m$ ellipses, that is, if and only if~$\sin\alpha$ is rational
\begin{eqnarray}
	\sin\alpha = \frac{m}{n} , \label{7.8}
\end{eqnarray}
with~$m$ and~$n$ relatively prime numbers and~$m \leq n$~\cite[p.~91]{GOLDSTEINhCla}. 

Furthermore, the numbers~$m$ and~$n$ define the topology of the orbits in terms of number of double points~\cite[p.~28]{dosSANTOSrMon}, classifying them into families of the same topology.

\section{Dyon-dyon scattering}
After the dyon-dyon bound states, seen in the previous section, let us now study the classical scattering~($\varepsilon>1$) of one dyon by another.

For this study, we need to obtain the so-called \emph{cross-section for scattering in a given direction} $\sigma(\Omega)$ and to arrive at it we will follow Goldstein's procedure~\cite[sec.~3.10]{GOLDSTEINhCla}.

From~\eref{6.20} and a reasoning similar to the one that leaded to~\eref{7.4}, this corresponds to
\begin{eqnarray}
	E > 0 , 
\end{eqnarray}
which, from~\eref{6.14}, analogously to~\eref{7.5}, implies a repulsive interaction, i.e.
\begin{eqnarray}
	q > 0 .
\end{eqnarray}

To start with, let us define~$ v_0 $ as the velocity of the dyon when it is at an infinite distance ($ r \to \infty $) from the other. Then, from~\eref{6.14}, the conserved energy reduces to the kinetic energy at that point:
\begin{eqnarray}
	E = \frac{1}{2}\mu v_0^2 . \label{8.1} 
\end{eqnarray}

We can now define~$ d $ as the distance of closest approach (periapsis) and impose the following condition on~$ \beta' $
\begin{eqnarray}
	\beta' = \beta(r = d) = 0 
\end{eqnarray}
into the general equation of the orbit~\eref{6.18}, obtaining
\begin{eqnarray}
	\frac{1}{r} = -\frac{\mu q}{4 \pi \epsilon_0 L^2}\left[1 + \varepsilon\cos\left( \beta\sin\alpha \right) \right] \label{8.2} 
\end{eqnarray}
with $\varepsilon$ now given by substituting~\eref{8.1} into~\eref{6.20} as
\begin{eqnarray}
	\varepsilon &= \sqrt{ 1 + 32 \pi^2 \epsilon_0^2 (\mu v_0^2 / 2) L^2 / \mu q^2 }  \nonumber \\
	&= \sqrt{1 + (4 \pi \epsilon_0 )^2L^2v_0^2/q^2} , \label{8.2b}
\end{eqnarray}
a result similar to that obtained by Schwinger~\cite{SCHWINGERjDyo}.

We can now study the dyon-dyon scattering by defining the scattering angle~$\Theta$, the angle between incidence~$\hat{\bi{r}}_i$ and scattering~$\hat{\bi{r}}_f$ directions at an infinite distance from the origin, as
\begin{eqnarray}
	\cos\Theta = -\hat{\bi{r}}_i \cdot \hat{\bi{r}}_f . \label{8.3}
\end{eqnarray}

Expressing~$\hat{\bi{r}}$ in the coordinate system ($r$, $\alpha$, $\beta$) by
\begin{eqnarray}
	\hat{\bi{r}} = \sin\alpha\:\cos\beta\:\hat{\bimath} + \sin\alpha\:\sin\beta\:\hat{\bjmath} + \cos\alpha\:\hat{\bi{k}} \label{8.4}
\end{eqnarray}
with~$\alpha$ constant and given by~\eref{6.9}, we obtain, from~\eref{8.3}, 
\begin{eqnarray}
	\cos\Theta &= - [ \sin^2\alpha ( \cos\beta_i\:\cos\beta_f + \sin\beta_i\:\sin\beta_f ) + \cos^2\alpha ] , \nonumber
\end{eqnarray}
where~$\beta_i$ and~$\beta_f$ are the azimuthal angles of the directions of incidence and scattering at an infinite distance from the origin, respectively, and, by making use of a few trigonometric identities such as
\numparts
 \begin{eqnarray}
	\cos^2\theta = 1 - \sin^2\theta  \label{trig1} \\
	\cos(\theta_1 - \theta_2) = \cos\theta_1\:\cos\theta_2 + \sin\theta_1\:\cos\theta_2 \label{trig2} \\
	\sin^2(\theta/2) = \frac{1 - \cos\theta}{2} \label{trig3} \\
	\cos(\theta/2) = \pm \sqrt{ (1 + \cos\theta) / 2 } \label{trig4} 
 \end{eqnarray}
\endnumparts
we obtain
\begin{eqnarray}
	\cos\Theta &= - [ \sin^2\alpha\:\cos(\beta_i\ - \beta_f) + (1 - \sin^2\alpha) ] \ \ \  \text{by~\eref{trig1} and~\eref{trig2} }   \nonumber \\ 
	&= \sin^2\alpha\:[1 - \cos(\beta_i\ - \beta_f)] - 1 \nonumber \\
	&= \sin^2\alpha\:\left\{2\sin^2[(\beta_i - \beta_f)/2]\right\} - 1  \ \ \  \text{by~\eref{trig3} } \nonumber \\
	&= 2\left\{ \sin\alpha\:\sin[(\beta_i - \beta_f)/2] \right\}^2 - 1 , \nonumber 
\end{eqnarray}
and, finally, by~\eref{trig4},
\begin{eqnarray}
	\cos(\Theta/2) = \sin\alpha\left|\sin\left[(\beta_i - \beta_f)/2\right]\right| . \label{8.5}
\end{eqnarray}

To proceed further, we need to calculate~$\beta_i$ and~$\beta_f$, the azimuthal angles of the directions of incidence and scattering at an infinite distance from the origin. Notice that $(1/r)_{r\to\infty} = 0$ and, therefore, from~\eref{8.2}, we get
\begin{eqnarray}
  \left( -\frac{\mu q}{4 \pi \epsilon_0 L^2}\left[1 + \varepsilon\cos\left( \beta\sin\alpha \right) \right] \right)_{r\to\infty} = 0 \nonumber
\end{eqnarray}
and we can evaluate~$\cos(\beta \sin\alpha)_{r\to\infty}$ as 
\begin{eqnarray}
  \cos(\beta \sin\alpha)_{r\to\infty} &= -1/\varepsilon \nonumber \\
	   &= - \left[ 1 + (4 \pi \epsilon_0 )^2 L^2 v_0 ^2 / q^2 \right]^{-1/2} \nonumber \\
	   &= - \left[ 1 + (4 \pi \epsilon_0 )^2 (\mu_0 \kappa \tan\alpha / 4 \pi)^2 v_0 ^2 / q^2 \right]^{-1/2} \nonumber \\
	   &= - \left( 1 + (\kappa^2 v_0^2 / c^4 q^2) \tan^2\alpha \right)^{-1/2}  \label{8.6}
\end{eqnarray}
where we used~\eref{8.2b},~\eref{6.8}, and~$ \mu_0\epsilon_0 = 1/c^2 $. 

Now, by making use of the trigonometric identity
\begin{eqnarray}
 \cos(\arctan x) = \frac{1}{\sqrt{1 + x^2}} , \nonumber
\end{eqnarray}
into~\eref{8.6}, we obtain
\begin{eqnarray}
 (\beta\sin\alpha)_{r\to\infty} &= \arctan{[( \kappa v_0 / c^2 |q| )\tan\alpha]} \nonumber \\
 &= \arctan{( k \tan\alpha)} , \label{8.7} 
\end{eqnarray}
where we introduced the parameter $k$ defined as
\begin{eqnarray}
 k \equiv \kappa v_0 / c^2 |q| . \label{k}
\end{eqnarray}

Therefore, from~\eref{8.5}, we finally obtain the scattering angle~$\Theta$ as
\begin{eqnarray}
	\cos(\Theta/2) = \sin\alpha\left|\sin(\Phi/\sin\alpha)\right| , \label{8.9}
\end{eqnarray}
where, for compactness, we introduced the parameter~$\Phi$ defined by
\begin{eqnarray}
 \Phi &\equiv (\beta\sin\alpha)_{r\to\infty} \nonumber
 &= \arctan{( k \tan\alpha)} , \label{8.8}
\end{eqnarray}
from the result~\eref{8.7}.

\begin{figure}[ht]
    \centering
    \includegraphics[width=0.8\textwidth]{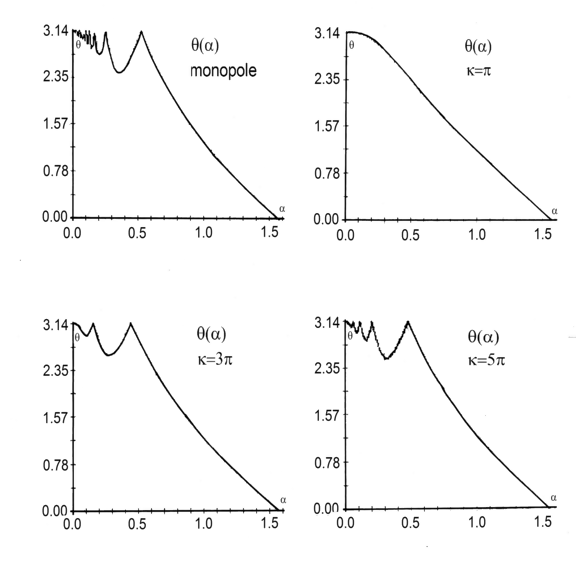}
    \caption{Graphs of the functional dependence of the scattering angle~$\Theta$ upon the half-aperture angle of the cone,~$\alpha$, according to~\eref{8.9}, for the monopole-electron pair and for various values of the parameter~$k \equiv \kappa v_0 / c^2 q$.}
    \label{fig_3}
\end{figure}

We present in \Fref{fig_3} graphics of the functional dependence of~$\Theta$ upon~$\alpha$ for various values of the parameter~$k$. In this figure, it is also displayed a graph for monopole-electron scattering, obtained by making~$e_2=g_1=0$ (which, from~\eref{5.2}, results in~$q=0$ and~$\kappa=-eg/c$) in~\eref{8.7}.

As the scattering orbit is asymptotic to~$\hat{\bi{r}}_i$ and~$\hat{\bi{r}}_f$, which define a plane, we can calculate the elastic differential cross-section~\cite[p.~108]{GOLDSTEINhCla} for the dyon-dyon scattering, in terms of the \emph{impact parameter}~$s$ (the perpendicular distance between the centre of force and the incident velocity)~\cite[p.~107]{GOLDSTEINhCla}, by the equation
\begin{eqnarray}
	\frac{\rmd \sigma}{\rmd \Omega}  = \left|\frac{s\rmd s}{d(\cos\Theta)}\right| = \sum_{s} s \left|\frac{\rmd(\cos\Theta)}{\rmd s}\right|^{-1} \label{8.10}
\end{eqnarray}
since, as seen from the graphs in \Fref{fig_3}, in general,~$\alpha$ is a multivalued function of~$\Theta$~\cite[p.~111]{GOLDSTEINhCla}. 

As $v_0$ is the incident velocity and perpendicular to the impact parameter~$s$, we can express the angular momentum~$L$, from~\eref{6.1}, as~$L=\mu sv_0$ and also substitute~\eref{6.8}, resulting   
\begin{eqnarray}
	s &= L / (\mu v_0) \nonumber \\
	 &= ( \mu_0 \kappa / 4 \pi \mu v_0 ) \tan\alpha , \label{8.10b}
\end{eqnarray}

Now, using $\rmd f(\alpha) / \rmd s = ( \rmd \alpha / \rmd s ) \rmd f(\alpha) / \rmd \alpha = ( \rmd  s/ \rmd \alpha )^{-1} \rmd f(\alpha) / \rmd \alpha $ and $\tan'\theta = \sec^2\theta = \cos^{-2}\theta $, we can rewrite~\eref{8.10b} as a relation between $\rmd s$ and $\rmd \alpha$ as
\begin{eqnarray}
  \frac{\rmd}{\rmd s} = \left[ \left(\frac{\mu_0 \kappa}{4 \pi \mu v_0}\right) \frac{1}{\cos^2\alpha} \right]^{-1} \frac{\rmd}{\rmd \alpha} ,
\end{eqnarray}
and use this and~\eref{8.10b} to evaluate the derivatives with respect to~$s$ in~\eref{8.10}, obtaining the elastic differential cross-section as 
\begin{eqnarray}
	\frac{\rmd \sigma}{\rmd \Omega} &= \sum_{\alpha} \left[ \left( \frac{\mu_0 \kappa}{4 \pi \mu v_0} \right) \frac{1}{\cos^2\alpha} \right] \left| \frac{\rmd(\cos\Theta)}{\rmd \alpha} \right|^{-1} \left[ \left(\frac{\mu_0 \kappa}{4 \pi \mu v_0}\right) \tan\alpha \right]  \nonumber \\
	&= \sum_{\alpha} \left( \frac{\mu_0 \kappa}{4 \pi \mu v_0} \right)^2 \frac{\sin\alpha}{\cos^3\alpha} \left| \frac{\rmd(\cos\Theta)}{\rmd \alpha} \right|^{-1} \nonumber \\
	&= \sum_{\alpha} \left(\frac{\mu_0 \kappa}{4 \pi \mu v_0}\right)^2 \frac{1}{2\cos^4\alpha} \left| \frac{\sin(2\alpha)}{\sin\Theta}\frac{\rmd \alpha}{\rmd \Theta} \right| , \label{8.11}
\end{eqnarray}
where, in the last step,  we used the trigonometric identity $\sin(2\alpha) = 2\sin\alpha\:\cos\alpha$.

Particularly interesting are the cases where the parameter~$k$ defined in~\eref{k} is a multiple of~$\pi$. One can see that by expanding~\eref{8.8} with~\eref{8.7} around~$\alpha=0$, using the approximations $\tan\theta \cong \sin\theta$ and $\arctan\theta \cong \theta$, as 
\begin{eqnarray}
	\Phi &= \arctan{ [ ( \kappa v_0 / c^2 |q| ) \tan\alpha ] } \nonumber \\
	 &\cong ( \kappa v_0 / c^2 |q| ) \sin\alpha , 
\end{eqnarray}
and using this result in~\eref{8.9}, obtaining
\begin{eqnarray}
	\cos(\Theta/2) &\cong \sin\alpha \left| \sin\{ [ ( \kappa v_0 / c^2 |q| ) \sin\alpha ] /\sin\alpha \} \right| \nonumber \\ 
	&\cong \alpha \left| \sin( \kappa v_0 / c^2 |q| ) \right| , \nonumber 
\end{eqnarray}
from which, by means of the trigonometric identity $ \cos(2\theta) = 2\cos^2\theta - 1 $, we obtain
\begin{eqnarray}
		\cos(\Theta) &\cong 2\alpha^2\sin^2( \kappa v_0 / c^2 |q| ) - 1 . 
\end{eqnarray}

With these approximations and the usual $ \sin\theta \cong \theta $ and $ \cos\theta \cong 1 $ ones, the cross-section~\eref{8.11} results  
\begin{eqnarray}
	\frac{\rmd \sigma}{\rmd \Omega} &\stackrel{\alpha \to 0}{\rightarrow} \sum_{\alpha} \left( \frac{\mu_0 \kappa}{4 \pi \mu v_0} \right)^2 \alpha \left| \frac{\rmd[ 2\alpha^2\:\sin^2( \kappa v_0 / c^2 |q| ) - 1 ]}{\rmd \alpha} \right|^{-1} \nonumber \\ 
	&\stackrel{\alpha \to 0}{\rightarrow} \sum_{\alpha} \left( \frac{\mu_0 \kappa}{4 \pi \mu v_0} \right)^2 \alpha \left| 4\alpha\: \sin^2( \kappa v_0 / c^2 |q| ) \right|^{-1} \nonumber \\
	&\stackrel{\alpha \to 0}{\rightarrow} \left(\frac{\mu_0 \kappa}{8 \pi \mu v_0}\right)^2 \frac{1}{\sin^2( \kappa v_0 / c^2 |q| )}  \label{8.32}
\end{eqnarray}
and, therefore, when~$\alpha \to 0$, a second order pole occur in the cross-section whenever
\begin{eqnarray}
 k = n\pi \qquad & (n=1,2,3,\cdots) . \label{8.33}
\end{eqnarray}

To evaluate the differential cross-section~\eref{8.11}, we found it convenient to introduce the new variable
\begin{eqnarray}
	\zeta &\equiv 2\Phi / \sin\alpha , \label{8.12}
\end{eqnarray}
where~$\Phi$ is given by~\eref{8.8} and~\eref{8.7}. 

Now, from~\eref{8.9} and again by means of the trigonometric identity $ \cos(2\theta) = 2\cos^2\theta - 1 $, we have
\begin{eqnarray}
	\cos(\Theta) = 2\sin\alpha^2\:\sin^2(\zeta/2) - 1 .
\end{eqnarray}

From this result, we can evaluate the derivative on the right-hand side of~\eref{8.11} as
\begin{eqnarray}
  \left(\frac{1}{\sin\Theta}\frac{\rmd \alpha}{\rmd \Theta} \right)^{-1} &= \frac{\rmd \cos\Theta}{\rmd \alpha} \nonumber \\
	&= \sin(2\alpha)\left[( 1 - \cos\zeta ) + \sin\zeta\tan\alpha\frac{\rmd(\zeta/2)}{\rmd \alpha} \right] . \label{8.13}
\end{eqnarray}

Then, from~\eref{8.12} and~\eref{8.8}, we can evaluate the derivative of~$\zeta$/2 as
\begin{eqnarray}
  \frac{\rmd (\zeta/2)}{\rmd \alpha} = \frac{\kappa v_0}{q\sin\alpha} \frac{\cos^2[(\zeta/2)\sin\alpha]}{\cos^2\alpha} - (\zeta/2)\cot\alpha \label{8.14}
\end{eqnarray}
and obtain, from~\eref{8.11}, the differential cross-section as 
\begin{eqnarray}
\frac{\rmd \sigma}{\rmd \Omega} = \left(\frac{\mu_0 \kappa}{4 \pi \mu v_0}\right)^2 g(\zeta) , \label{8.16}
\end{eqnarray}
where
\begin{eqnarray}
g(\zeta) = \sum_{\alpha}\frac{1}{\cos^4\alpha} \frac{1}{ \left| 2(1-\cos\zeta) - \zeta\sin\zeta + (\sin\zeta/\sin\alpha)\sin(\zeta\sin\alpha) \right| } , \label{8.17}
\end{eqnarray}
with~$\zeta$ given by~\eref{8.12}.

\begin{figure}[ht]
    \centering
    \includegraphics[width=0.8\textwidth]{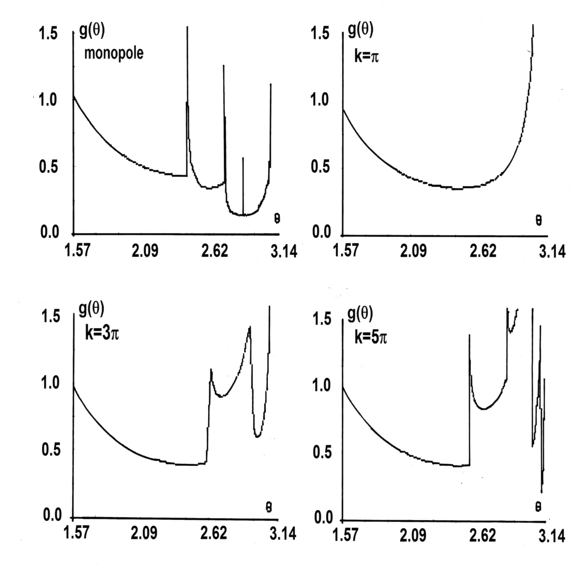}
    \caption{Differential cross-section~$g(\Theta)$ scattering, according to~\eref{8.17}, for the monopole-electron pair and for various values of the parameter~$k \equiv \kappa v_0 / c^2 |q|$.}
    \label{fig_4}
\end{figure}

In \Fref{fig_4}, we present graphs of the differential cross-section for various values of the parameter~$k$. In this figure, we also show a graph for the electron-monopole scattering obtained, again, by making~$q=0$ and~$\kappa=-eg$ in~\eref{8.17}.

One observes from \Fref{fig_4} that the cross-section becomes infinite for some values of~$\Theta$ and falls abruptly thereafter. From~\eref{8.11}, we see that it happens, besides~$\Theta = 0$, in one of the following cases
\begin{equation}
  \begin{cases}
    \rmd \Theta/\rmd \alpha = 0 &  \\  
    \Theta=\pi & \mathrm{but } \: \sin(2\alpha)\neq 0 .
  \end{cases} 
	\label{8.18}
\end{equation}

This phenomenon is very similar to what occurs in the optical scattering of sunlight by raindrops and, because of this similarity, these two conditions are referred to as \emph{rainbow scattering}~\cite[p.~111]{GOLDSTEINhCla} and \emph{glory scattering}~\cite[p.~ 114]{GOLDSTEINhCla}, respectively. 

In the case of rainbow scattering, we have, from~\eref{8.9},
\begin{eqnarray}
	\tan\left(\frac{\Phi_\mathrm{r}}{\sin\alpha_\mathrm{r}}\right) - \frac{\Phi_\mathrm{r}}{\sin\alpha_\mathrm{r}} + \frac{\sin(2\Phi_\mathrm{r})}{\sin(2\alpha_\mathrm{r})\cos\alpha_\mathrm{r}} = 0 , \label{8.23}
\end{eqnarray}
where the suffix~$_\mathrm{r}$ was appended to indicate that it refers to the rainbow scattering. \Eref{8.23} can be solved numerically. In \Tref{tab:rainbow}, we present rainbow~$\alpha_\mathrm{r}$ angles for two values of the~$k$ parameter defined in~\eref{k}.

\Table{\label{tab:rainbow}Rainbow angles~$\alpha_\mathrm{r}$ ($\rmd \Theta/\rmd \alpha=0$) obtained from~\eref{8.23}.}
			\br
			$k$ & $\alpha_\mathrm{r}$ \\
			\mr
			$2\pi$ & 0.230 \\
			\mr
			$3\pi$ & 0.104 \\
   			& 0.280 \\
			\br
\endTable

For the glory scattering, we have, from~\eref{8.9},
\begin{eqnarray}
\Phi_\mathrm{g}/\sin\alpha_\mathrm{g} = n\pi \qquad & (n=1,2,3,\cdots) , \label{8.19}
\end{eqnarray}
and, from~\eref{8.8}, the condition
\begin{eqnarray}
\tan(n\pi\sin\alpha_\mathrm{g}) = k\tan\alpha_\mathrm{g} \qquad & (n=1,2,3,\cdots) , \label{8.20}
\end{eqnarray}
which can be solved numerically for the desired value of the parameter~$k$. In \Tref{tab:glory}, we present glory~$\alpha_\mathrm{g}$ angles for two values of the parameter~$k$.
 
\Table{\label{tab:glory}Glory angles~$\alpha_\mathrm{g}$ ($\Theta=\pi$) obtained from~\eref{8.20}.}
			\br
			$k$ & $\alpha_\mathrm{g}$\\
			\mr
			$2\pi$ & 0.394 \\
			\mr
			$3\pi$ & 0.156 \\
					& 0.445 \\
			\br
\endTable

Let us now consider what happens at small-angle scattering~($\alpha \to \pi/2$). From~\eref{8.9} and using the trigonometric identity $ \sin\arctan\theta = x / \sqrt{ 1 + x^2 } = 1 / \sqrt{ 1 + x^{-2} } $,
 we have
\begin{eqnarray}
\cos(\Theta/2) &= \sin\alpha \left| \sin [ \arctan(k\tan\alpha) ] \right| \nonumber \\
 &= \frac{\sin\alpha}{\sqrt{ 1 + (k\tan\alpha)^{-2} }} . \label{8.28}
\end{eqnarray}

Now, using~\eref{6.8} and expressing the parameter~$k$ in terms of the impact parameter~$s$ by means of~\eref{8.10b}, we have
\begin{eqnarray}
	\tan\alpha &= (4\pi / \mu_0\kappa) L  \nonumber \\
   &= (4\pi / \mu_0\kappa) (\mu v_0 s)  \nonumber \\
   &= (4\pi\mu v_0 / \mu_0\kappa) s , \label{8.28b}
\end{eqnarray}
and, by means of the trigonometric identity $ \sin\theta = 1 / \sqrt{ 1 + \cot^2\theta } $, we obtain
\begin{eqnarray}
	\sin\alpha = 1 / \sqrt{ 1 + (\mu_0\kappa / 4\pi\mu v_0 s)^2 } .  \label{8.28c}
\end{eqnarray}

On the other hand, from~\eref{k} and~\eref{8.28b}, we get
\begin{eqnarray}
	k\tan\alpha &= (\kappa v_0 / c^2 |q|) ( 4\pi\mu v_0 / \mu_0\kappa ) s \nonumber \\
  &= ( 4\pi\epsilon_0\mu v_0^2 s / |q| ) ,
\end{eqnarray}
where we used~$ \mu_0\epsilon_0 = 1/c^2 $ once more, and, therefore, 
\begin{eqnarray}
	\frac{1}{ \sqrt{ 1 + (k\tan\alpha)^{-2} } } = \frac{1}{ \sqrt{ 1 + (q / 4\pi\epsilon_0\mu v_0^2 s)^2 } } .  \label{8.28d}
\end{eqnarray}

Substituting~\eref{8.28c} and~\eref{8.28d} into~\eref{8.28}, using the approximation $ 1/\sqrt{1+x} \cong 1 - x/2 $, and keeping terms only up to first order, we obtain
\begin{eqnarray}
	\cos(\Theta/2) &= \frac{1}{ \sqrt{ 1 + \left( q / 4 \pi \epsilon_0 \mu v_0^2 s \right)^2 } }  \frac{1}{ \sqrt{ 1 + \left( \mu_0 \kappa / 4 \pi \mu v_0 s \right)^2 } } \nonumber \\
	&\cong \left[ 1 - \frac{1}{2} \left( \frac{q}{4 \pi \epsilon_0 \mu v_0^2 s} \right)^2 \right] \left[ 1 - \frac{1}{2} \left( \frac{\mu_0 \kappa}{4 \pi \mu v_0 s}\right)^2 \right] \nonumber \\
	&\cong 1 - \frac{1}{2} \left[ \left( \frac{q}{4 \pi \epsilon_0 \mu v_0^2 s} \right)^2 + \left( \frac{\mu_0 \kappa}{4 \pi \mu v_0 s}\right)^2 \right] \nonumber \\
	&\cong 1 - \frac{1}{2} \frac{1}{(4 \pi \epsilon_0 \mu v_0)^2 s^2} \left[\left(\frac{q}{v_0}\right)^2 + \left( \frac{\kappa}{c^2} \right)^2 \right] \label{8.29a} .
\end{eqnarray}

Now, considering the approximation $ \cos(\Theta/2) \cong 1 - (\Theta/2)^2/2 $ and comparing this expression with~\eref{8.29a}, we have
\begin{eqnarray}
	(\Theta/2)^2 &\cong \frac{1}{(4 \pi \epsilon_0 \mu v_0)^2 s^2} \left[\left(\frac{q}{v_0}\right)^2 + \left( \frac{\kappa}{c^2} \right)^2 \right]
\end{eqnarray}
or
\begin{eqnarray}
	s^2 &\cong \frac{1}{(\Theta/2)^2} \frac{1}{(4 \pi \epsilon_0 \mu v_0)^2 } \left[\left(\frac{q}{v_0}\right)^2 + \left( \frac{\kappa}{c^2} \right)^2 \right] . \label{8.30}
\end{eqnarray}

Thus, from~\eref{8.10} and the approximation $\sin\theta \cong \theta$, we can compute the differential cross-section of the dyon-dyon scattering at small angles as
\begin{eqnarray}
	\frac{\rmd \sigma}{\rmd \Omega} &= s \left| \frac{\rmd s}{\rmd(\cos\Theta)} \right| \nonumber \\
	&= s \left| \frac{\rmd s}{\rmd (s^2)} \right| \left| \frac{\rmd (s^2)}{\rmd [(\Theta/2)^2]} \right| \left| \frac{\rmd [(\Theta/2)^2]}{\rmd (\Theta/2)} \right| \left| \frac{\rmd (\Theta/2)}{\rmd (\Theta)} \right| \left| \frac{\rmd (\Theta)}{\rmd (\cos\Theta)} \right| \nonumber \\
	&= s \left| \frac{\rmd (s^2)}{\rmd s} \right|^{-1} \left| \frac{\rmd [(\Theta/2)^2]}{\rmd (\Theta/2)} \right| \left| \frac{\rmd (\Theta/2)}{\rmd (\Theta)}\right| \left| \frac{\rmd (\cos\Theta)}{\rmd (\Theta)} \right|^{-1}  \left| \frac{\rmd (s^2)}{\rmd [(\Theta/2)^2]} \right| \nonumber \\
	&= s |2s|^{-1} \left| 2(\Theta/2) \right| \frac{1}{2} \left| -\sin\Theta \right|^{-1} \left| \frac{\rmd (s^2)}{\rmd [(\Theta/2)^2]} \right| \nonumber \\
	&\cong \frac{1}{4} \frac{|\Theta|}{|\sin\Theta|} \frac{1}{(4 \pi \epsilon_0 \mu v_0)^2}  \left[ \left( \frac{q}{v_0}\right)^2 + \left( \frac{\kappa}{c^2} \right)^2 \right] \left| \frac{\rmd }{\rmd [(\Theta/2)^2]} \left( \frac{1}{(\Theta/2)^2} \right) \right| \nonumber \\
	&\cong \frac{1}{4} \frac{1}{(4 \pi \epsilon_0 \mu v_0)^2}  \left[ \left( \frac{q}{v_0}\right)^2 + \left( \frac{\kappa}{c^2} \right)^2 \right] \left| \frac{-1}{(\Theta/2)^4} \right| \nonumber \\ 
	&\cong \frac{1}{4} \frac{1}{(4 \pi \epsilon_0 \mu v_0)^2}  \left[ \left( \frac{q}{v_0}\right)^2 + \left( \frac{\kappa}{c^2} \right)^2 \right] \csc^4(\Theta/2) \label{8.31}
\end{eqnarray}
which is a generalization of the Rutherford formula for the scattering of~$\alpha$ particles by atomic nuclei~\cite[p.~110]{GOLDSTEINhCla}.

A relativistic extension of the entire calculation previously done is possible, in a simple way~\cite[cap.~4]{dosSANTOSrMon}. The importance of a relativistic extension for the treatment of dyon-dyon system stems from the large value of the coupling constant for magnetic charges (see~\eref{2.10}). In the same way as in the non-relativistic case, the relativistic dyon is confined to the surface of a cone of half-aperture angle given by~\eref{6.8}. The bound states correspond to conic-shaped orbits confined to the surface of the Poincar\'e cone (\Fref{fig_2}), with a similar condition for the orbit closing. For the relativistic scattering case, the orbits are hyperbolas confined to the surface of the Poincar\'e cone and this system exhibits glory and rainbow scattering. 

It is even possible to apply the well-known Sommerfeld (semi-classical) quantization rule~\cite[p.~283]{SOMMERFELDaAto} to the dyon-dyon system. By doing it, we have shown the analogy between this system and the hydrogen atom and that the former can be considered its generalization~\cite[cap.~3]{dosSANTOSrMon}. Furthermore, we obtained, for both the non-relativistic and relativistic cases, an energy spectrum that reasonably approximates the quantum spectrum obtained by Pereira~\cite{PEREIRAjMon}. 

\section{Concluding remarks}
In this work, thanks to a Lagrangian without \emph{Dirac strings}, it was possible to make a classical study of the dyon-dyon system, with the electron-monopole system as a particular case.

The orbit equations were interpreted as representing conic-shaped orbits confined to the surface of the \emph{Poincar\'e cone}. While the electron does not form bound states with the magnetic monopole, the dyon-dyon system exhibits non-planar stable elliptic-like orbits, and we presented the conditions for them to be closed. 

Through the orbit equations, it was also possible to study the classical scattering for these systems. We showed that the elastic differential cross-section shows divergences that are usually denominated glory and rainbow scatterings by their similarities with Optics. We also showed that the differential cross-section of the dyon-dyon scattering at small angles is a generalization of the Rutherford formula.

Finally, we want to stress that a relativistic extension of this study can be done in a simple way, as well as a semi-classical quantization via Sommerfeld rule.

We hope that these results may arouse students' and teachers' interest and contribute to Classical Mechanics courses.

\ack
We would like to thank the late Prof. Paulo Leal Ferreira, who was my adviser through the master's degree at the Institute for Theoretical Physics in S\~ao Paulo, where the thesis, on which this work is based, was done. We also acknowledge FAPESP (S\~ao Paulo Research Foundation) for a scholarship that made the M.Sc. research work possible.

%62

\section*{References}

\bibliographystyle{iopart-num}

\bibliography{magneticmonopoles}

%306

\end{document}